\documentclass{emulateapj}
\slugcomment{Submitted to the Astronomical Journal}

\shorttitle{Small-Scale X-ray Variability of Cas A}
\shortauthors{Patnaude \& Fesen}

\begin{document}

\title{Small-Scale X-ray Variability in the Cassiopeia A Supernova
Remnant}

\author{Daniel J.~Patnaude\altaffilmark{1} \& Robert A.~Fesen 
\altaffilmark{2}}
\altaffiltext{1}{Smithsonian Astrophysical Observatory, 60 Garden St,
Cambridge, MA 02138}
\altaffiltext{2}{6127 Wilder Laboratory, Physics \& Astronomy
Department,
Dartmouth College, Hanover, NH 03755}

\begin{abstract}

A comparison of X-ray observations of the Cassiopeia A supernova remnant taken 
in 2000, 2002, and 2004 with the {\it Chandra} ACIS-S3 reveals the presence of 
several small scale features ($\leq 10\arcsec$) which exhibit significant 
intensity changes over a 4 year time frame. Here we report on the variability 
of six features, four of which show count rate increases from $\sim$ 10\% to 
over 90\%, and two which show decreases of $\sim$ 30\% -- 40\%. While 
extracted 1-4.5 keV X-ray spectra do not reveal gross changes in emission line 
strengths, spectral fits using non-equilibrium ionization, metal-rich plasma 
models indicate increased or decreased electron temperatures for 
features showing increasing or decreasing count rates, respectively. Based on the 
observed count rate changes and the assumption that the freely expanding 
ejecta has a velocity of $\sim$ 5000 km s$^{-1}$ at the reverse shock front, 
we estimate the unshocked ejecta to have spatial scale variations of $0.02 - 
0.03$ pc, which is consistent with the X-ray emitting ejecta belonging to a 
more diffuse component of the supernova ejecta than that seen in the optically 
emitting ejecta, which have spatial scales $\sim$ 10$^{-3}$ pc.

\end{abstract}
\section{Introduction}

Cassiopeia A (Cas A) is currently the youngest known Galactic supernova 
remnant (SNR) with an estimated explosion date no earlier than around 1670 AD 
\citep{thorstensen01,Fesen06a}.  The remnant consists of optical, infrared, and X-ray 
emission arranged in a shell roughly 4$\arcmin$ diameter ($\simeq$ 4 pc at 
3.4$_{-0.1}^{+0.3}$ kpc; \citealt{reed95}) consisting of undiluted supernova 
(SN) ejecta rich in O, Si, Ar, Ca, and Fe heated by the remnant's  $\simeq$ 
3000 km s$^{-1}$ reverse shock \citep{chevalier78,chevalier79,douvion99,
hughes00,willingale03,hwang03,laming03,morse04}.

Viewed in X-rays, the remnant's brightest emission is concentrated in a 30$
\arcsec$ thick, $\simeq$ 110$\arcsec$ radius emission ring dominated by thermal
plasma arising from shock heated SN debris. Farther out beyond this emission
ring, lies fainter filamentary X-ray emission (radius $\simeq$ 150$\arcsec$)
associated with the current position of the remnant's $\approx $ 6000 km
s$^{-1}$ forward blast wave as it moves through the surrounding circumstellar
material \citep{gotthelf01,delaney03}.

If the remnant's expanding SN ejecta were uniform in density and chemical 
composition, X-ray emission from the shocked ejecta would appear as a uniform 
ring of emission with little spectral variation. What is actually observed, 
however, is a patchy, irregular ring of clumpy emission reflecting the  
inhomogeneous nature of the expanding SN debris as it interacts with the 
reverse shock front. Due both to strong ejecta clumping caused by radiative 
cooling instabilities and the turbulent mixing of different chemical layers of 
the progenitor star during the SN explosion, the resulting X-ray emission 
morphology is structurally and chemically complex \citep {hughes00,
hwang00}.

Although Cas A's reverse shock front has been located to arcsecond accuracy in
a few regions through high-resolution optical images \citep{Fesenetal01,
morse04}, it has only been broadly localized through a sharp increase in X-ray
emission in azimuthal averages of selected portions of the remnant
\citep{gotthelf01}. On small spatial scales, evidence for the reverse shock
interacting with the ejecta would be manifested in X-rays as a rise in emission
as the SN ejecta is shocked and decelerated. Post-shock cooling of the
metal-rich ejecta could lead to equally dramatic localized emission declines. 

Here we present an 
analysis of archival {\sl Chandra} images of Cas~A which show a 
number of substantial, small-scale brightness changes between 2000 and 2004 
which we interpret as marking regions of reverse shock passage. We report on 
spectral analyses for six specific regions and find some plasma temperature 
changes in the X-ray emitting ejecta. In $\S$2 we present the relevant {\it 
Chandra} observations and subsequent data reduction. In $\S$3 we discuss the 
results from our analysis, and summarize our conclusions in $\S$4.

\section{Observations and Data Reduction}

Cas A was observed with the ACIS-S3 chip as part of the {\it Chandra} Guest 
Observer program for 50 ksec both on 30 January 2000 and 6 February 2002, and 
for 1 Msec as a {\it Chandra} Very--Large Project during the first half of 
2004 (see \citealt{hwang00,gotthelf01,delaney03} for details regarding the 
2000 and 2002 observations, and \citealt{hwang04} for the 2004 observations.). 
The $0\farcs492$ CCD pixel scale of ACIS undersamples the telescope's 
$\simeq 0 \farcs5$ resolution.  We reprocessed these data using the latest 
version (3.2) of the {\it Chandra} Calibration Database (CalDB). This 
reprocessing focused on recalculating the aspect solution for better 
positional accuracy. Version 3.2 of the CalDB was also used in the generation 
of detector response matrices used in the spectral fitting, which is 
discussed below.

\begin{figure}
\plotone{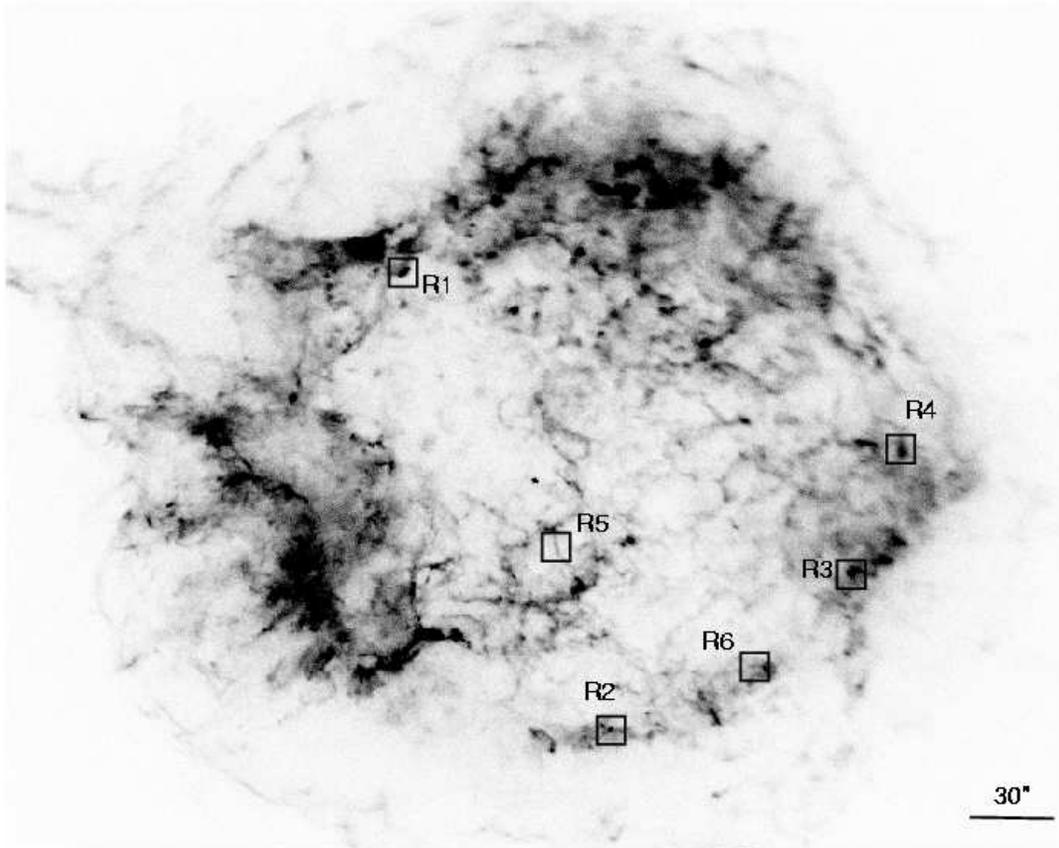}
\caption{A {\it Chandra} image of Cas A in 2004 (Observation ID
5196). The marked regions
correspond to those listed in Table~\ref{tab:rates} and shown in
Figure~\ref{fig:bright}.}
\label{fig:casa}
\end{figure}

Both direct visual comparisons of the three epoch images and image differences 
and subtractions were used to identify several small scale regions where the 
detected flux appeared to vary with time.  Six selected regions showing 
significant flux variations in the total ACIS band (0.3 -- 10.0 keV) are 
marked in Figure~\ref{fig:casa} with enlargements shown for the years 2000, 
2002, and 2004 in Figure~\ref{fig:bright}. The coordinates, sizes of the 
extracted flux regions, and count rates are given in Table~\ref{tab:rates}. 
The quoted rates for each epoch are normalized to the observed count rate of 
the remnant's central X-ray point source \citep{Tananbaum99,Aschen99,Pavlov99}.

Spectra were extracted for the six regions listed in Table~\ref{tab:rates}.
Background subtraction was done by selecting regions close to the knots. 
The background spectra were binned such that each detector channel contained
a minimum of 10 counts. The background spectra were then fit to models 
appropriate for a thermal plasma in non-equilibrium ionization. More
specifically, since we are investigating regions in the interior of the
SNR, the background contains contributions from swept up circumstellar
material as well as diffuse emission from shocked supernova ejecta around
the regions of interest. This
method of background subtraction is preferred to using ACIS Blank-Sky images
since scattered X-rays from the SNR are the dominant source of background here.
Furthermore, while the source spectra were found to vary, the background
spectra comes from the diffuse emission around the knots and does not vary. 

In order to minimize the effects of {\it Chandra's} azimuthally varying point
spread function which are manifest in the combined 2004 observations
\citep{hwang04},  we limited ourselves to using a 50 ksec subset (ObsID 5196; 8
February 2004) of the 1 Msec observation when extracting the 2004 spectra,
since the observational parameters for these data closely match those of the
2000 (ObsID 114) and 2002 (ObsID 1952) observations. The extracted source
and background spectra were
fit using version 11.2 of the X-ray spectral fitting package XSPEC.

The six regions of interest are relatively small in area ($\lesssim$ 30 
arcsec$^2$) with total observed counts of 5000 -- 20,000 for the 0.3 -- 10 keV 
energy a range for each of the three epoch observations.  We grouped the 
extracted spectra such that there was a minimum of 10 counts in any detector 
channel. Moreover, since the goal of this analysis was not concerned with 
looking for specific abundance variations in the ejecta, we chose to limit our 
spectral fitting to NEI models \citep{borkowski01} containing mainly Si, S, 
and Ar (and Mg and Ca, when there were prominent emissions visible at $\sim$ 
1.2 and 3.8 keV, respectively) and limited the fits to between 1.0 and 
4.5 keV. 
For all models, the ratio (Si/Si$_{\sun}$) was fixed at 1, while the relative 
abundances of the other included elements were allowed to vary. Finally, we 
chose to freeze the abundances of all other modeled elements (H, He, C, N, O, 
Ne, Ca, Fe, and Ni) at zero. We justify this by noting that these elements
do not have emission lines in the energy band we are interested in. We do
note, however, that by not including these elements, we may under-predict the 
contribution to the flux from the underlying continuum. In 
Table~\ref{tab:mfits}, 
we list the fitted plasma temperature for each epoch and each region, and show 
the data and the resulting fits in Figures~\ref{fig:reg12}--\ref{fig:reg56}.

All regions, with the exception of Region 5, are well modeled as a thermal 
spectrum. Region 5 appears to possess purely nonthermal emission and is 
likely filamentary emission arising from the remnant's forward shock front 
viewed in tangent like several other interior projected filamentary features 
(\citealt{hughes00}; Region `D') and common in the remnant's southwestern 
quadrant \citep{delaney04}.

\section{Results and Discussion}

\subsection{Count Rate Variations}

A comparison of 2000 -- 2004  {\it Chandra} images of Cas A shows fairly 
steady X-ray emission flux from the remnant's shock heated ejecta on both 
large and small spatial scales.  However, there are a few small features in 
the main emission ring that do show significant changes in brightness over 
this time interval. Most of these show increases rather than decreases 
although this may be due in part to a detection bias against finding changes 
in diffuse patches of emission, especially if they are faint.

\begin{figure}
\epsscale{0.90}
\plotone{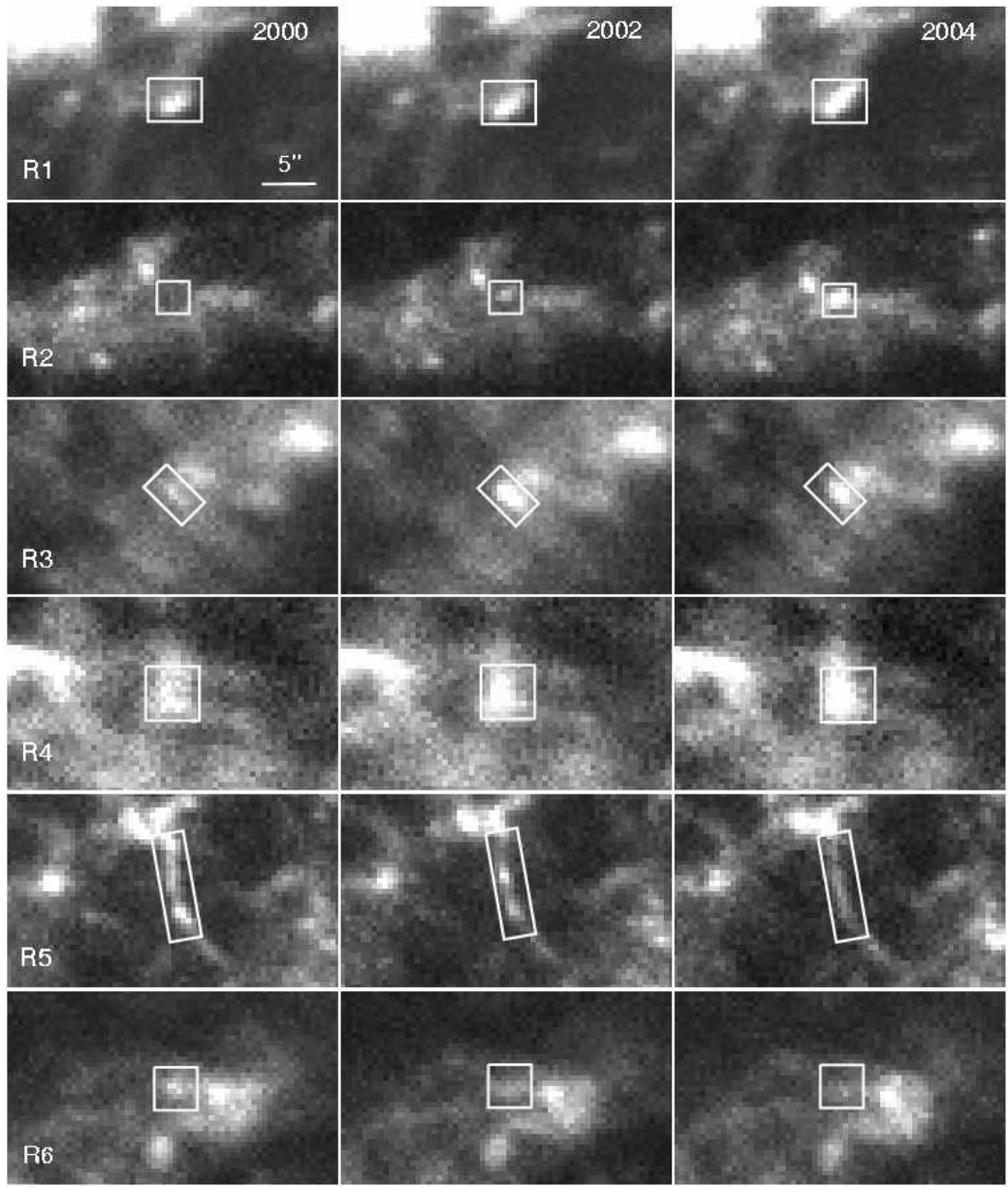}
\caption{Close-up of the six regions listed in Table~\ref{tab:rates}
for each
epoch. The boxes correspond to the spectral extraction
regions used in the fits in Table~\ref{tab:mfits} and shown in
Figures~\ref{fig:reg12}--\ref{fig:reg56}.}
\label{fig:bright}
\end{figure}

As can be seen qualitatively in Figure~\ref{fig:bright} and quantitatively in 
Table~\ref{tab:rates}, between 2000.1 and 2004.1 four small regions in Cas~A's 
X-ray emission structure exhibited significant increases in intensity while 
two 
regions showed a decrease. Specifically, Regions 1, 3, and 4 each show 
increases of $\approx$ 10\%--20\% between 2000 and 2004, while Region 2 went 
from near invisibility in 2000 to being among the  brightest 
features in the remnant, with a near doubling of its count rate between 2002 
and 2004. In 2000, the count rate in this small region is due 
mainly to the diffuse emission in the area around and toward the knot in 
Region 2. As can seen in the 2000 and 2002 images, a new and relatively bright knot 
emerges here, becoming twice as bright just two years later.

We note that fine-scale flux changes in the remnant are not necessarily 
monotonic. Region 3, for example, showed both an increase and a decrease over 
four years; nearly a 40\% increase from 2000 to 2002 but then a 14\% decrease 
from 2002 to 2004. The short time span of data taken in early 2004, taken as 
part of the 1Msec Very Large Program on Cas A, does not permit us to 
investigate whether this decrease continued after February 2004, but future 
observations could investigate this question.

Most of these flux changes occur on relatively small spatial scales, of order 
a few arcseconds. For example, the white box in Figure 2 centered on the 
emerging 
bright spot in Region 2 is just $3'' \times 3''$ and yet encloses most of the 
observed count rate change. Similarly, the majority of the brightness change 
for the northeastern limb emission knot shown in Region 1 occurs along a 
segment of the knot's structure extending 1'' - 3'' to the northwest.

In contrast to the brightening seen for Regions 1--4, the X-ray emission 
detected in Regions 5 and 6 show decreases in brightness. Shown in the bottom 
two rows of Figure~\ref{fig:bright}, the count rates in these regions drop 
significantly from 2002 to 2004; a $\sim$ 40\% decrease for Region 5 and 
$\sim$ 
20\% in Region 6.  For Region 5, the measured 40\% change in flux represents 
an 
average over several small, individual features some of which exhibited large 
and small changes in brightness.

While it has been known for some time that a layer of contamination is forming 
over the ACIS detector thus reducing its overall 
efficiency \citep{plucinsky03}, the expected flux reduction from this effect is not enough to 
explain the overall decreases we see here. Nonetheless, to avoid uncertainties 
due to detector sensitivity degradation, we list in Table 2 measured count 
rates normalized to the observed count rate for the remnant's central X-ray 
point source.  Conversion of the observed count rates into intensity  changes 
can be estimated using the X-ray point source observed $F_{X} = 8 \times 
10^{-13}$ erg cm$^{-2}$ (for $0.6-6$ keV; see \citealt{Fesen06b} and 
refs. therein).

\subsection{Spectral Variations}

\begin{figure}
\plottwo{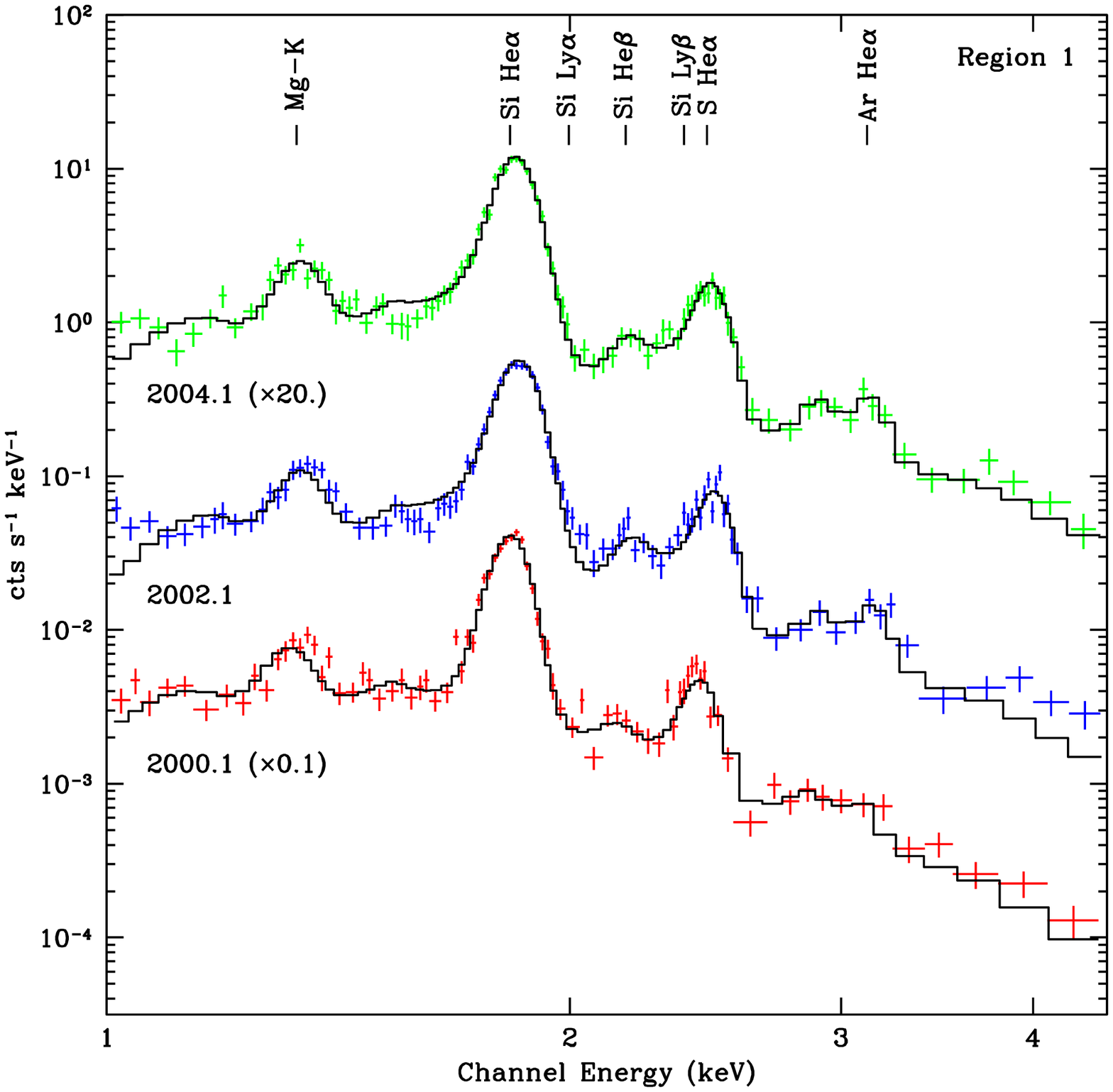}{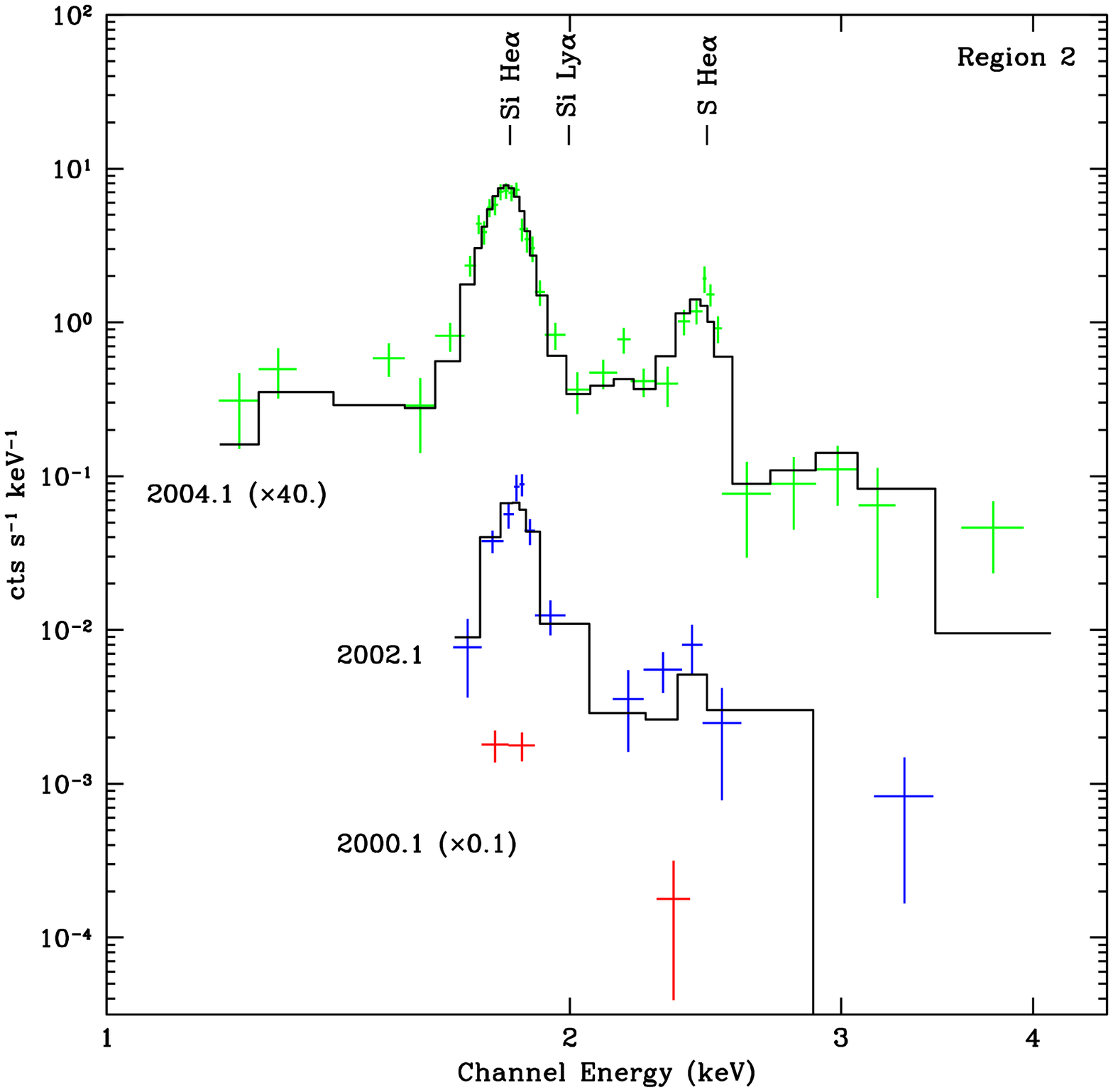}
\caption{{\it Left}: Spectral fits to Region 1 for 2000, 2002 and
2004. {\it Right}:
Region 2 data for 2000 and spectral fits to Region 2 for 2002 and 2004.}
\label{fig:reg12}
\end{figure}

\begin{figure}
\plottwo{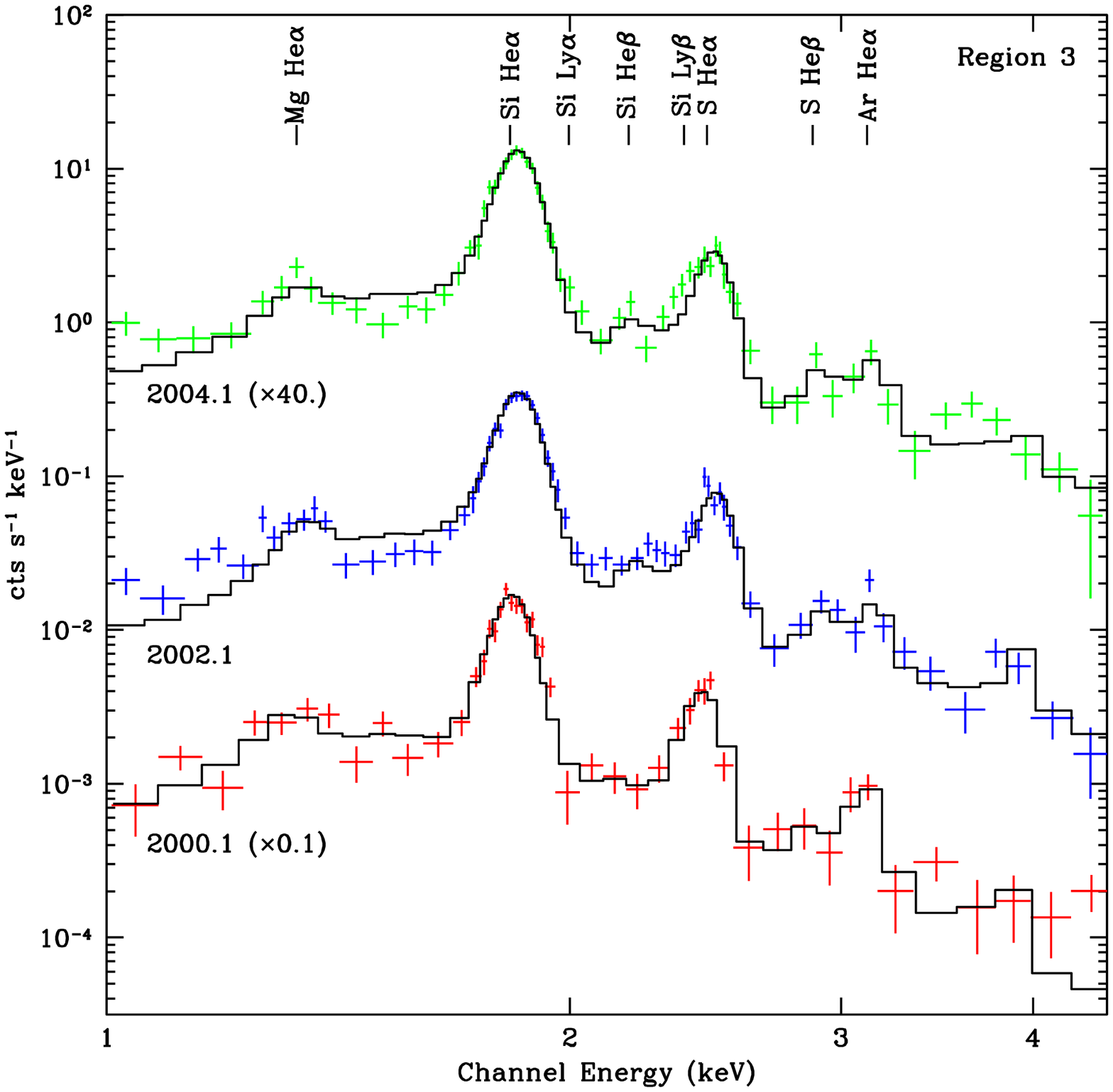}{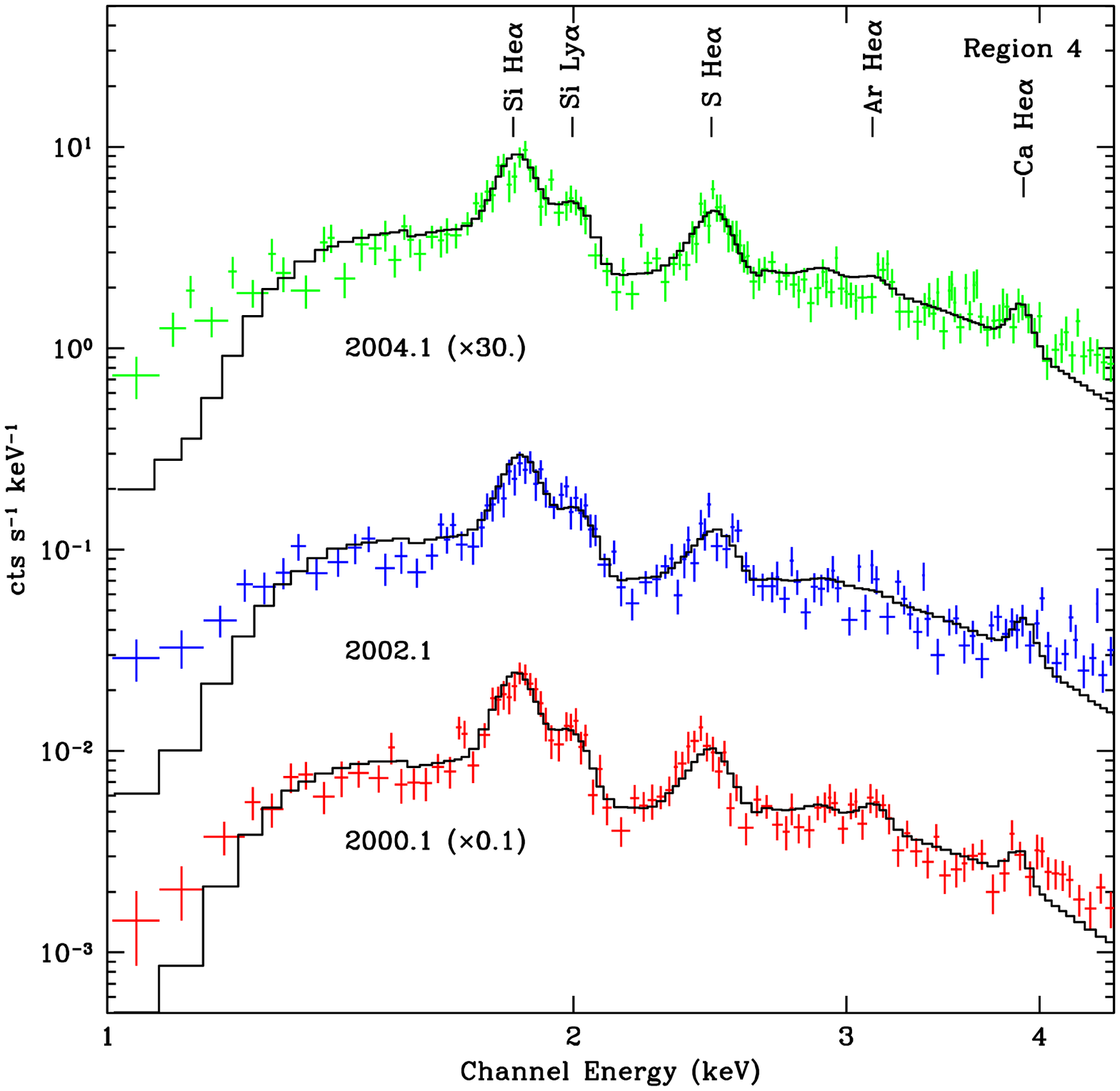}
\caption{{\it Left}: Spectral fits to Region 3 for 2000, 2002 and
2004. {\it Right}:
Same as {\it left}, for Region 4.}
\label{fig:reg34}
\end{figure}

\begin{figure}
\plottwo{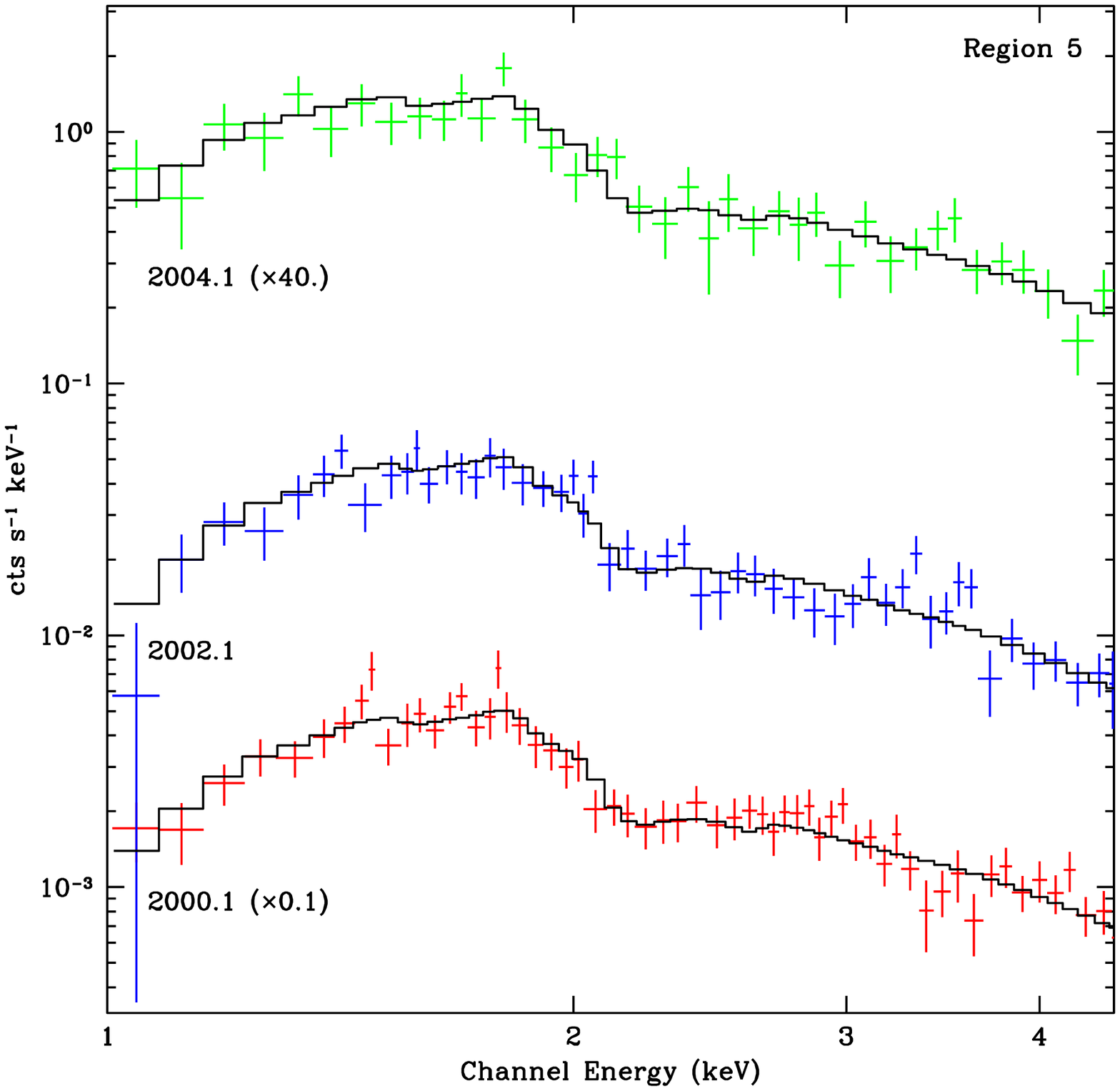}{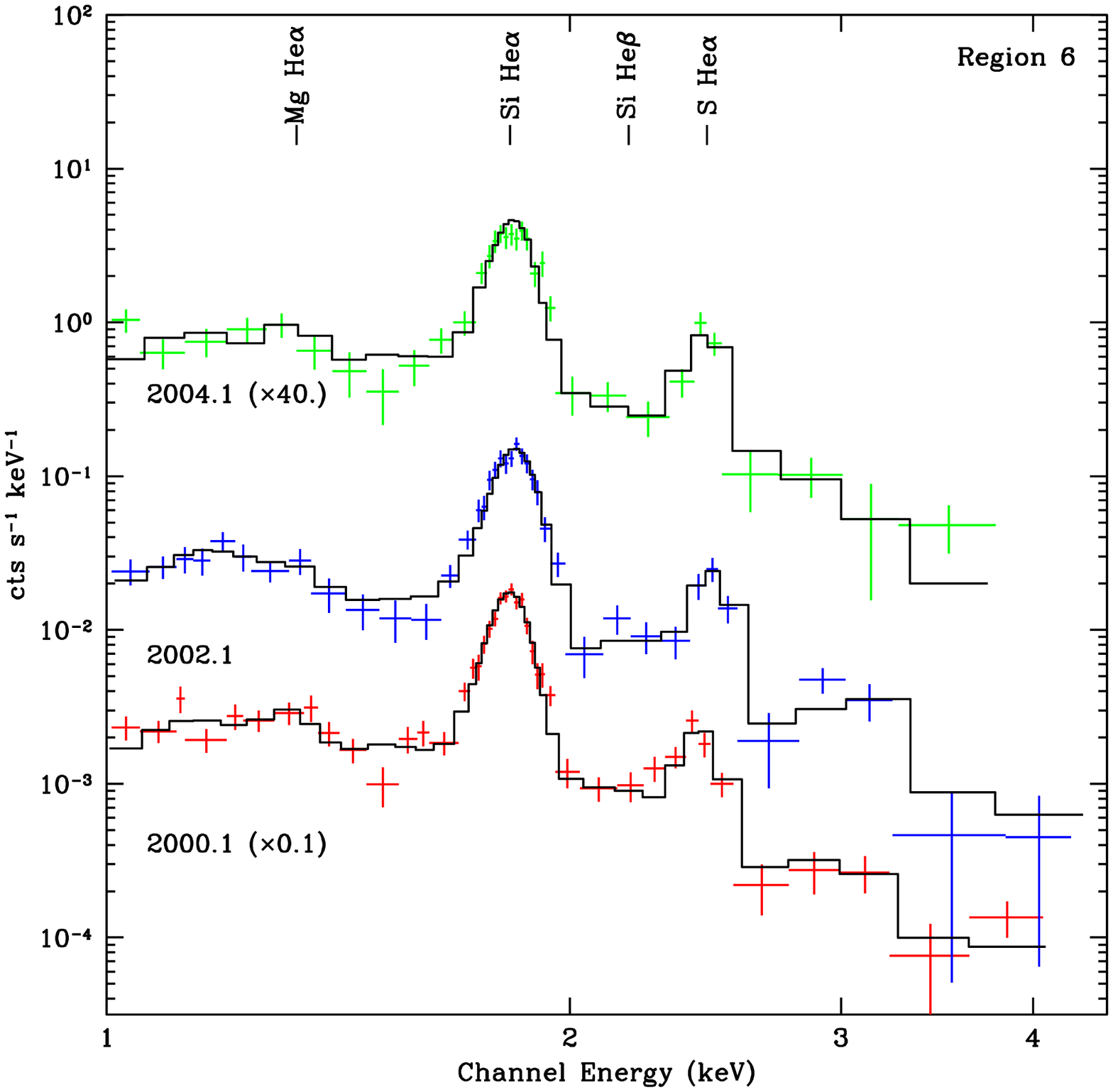}
\caption{{\it Left}: Spectral fits to Region 5 for 2000, 2002, and
2004. {\it Right}:
Same as {\it left} for Region 6.}
\label{fig:reg56}
\end{figure}

We performed spectral fits for each region to investigate whether rapid changes
in the observed count rate also reflected changes in spectral properties of the
emission features. In general, we find that the modeled spectra are relatively
constant from year to year (cf., Region 1, Fig.~\ref{fig:reg12}, left, and
Regions 4, Fig.~\ref{fig:reg34}, right), but there are exceptions. For example,
Region 2 (Fig.~\ref{fig:reg12}, right) which showed the most extreme flux
brightening, shows  some evidence for the emergence of helium-like silicon and
sulfur emission starting in the 2002 spectrum. From our spectral fits for this
region, we also found that $kT$ increased in value from $\sim$ 1.0 -- 1.5 keV
between 2002 and 2004, combined with an increase in $n_e t$ from 4.6 -- 53
$\times$ 10$^{10}$ cm$^{-3}$ s over the same time period. Such large increases
in flux and ionization timescale suggest that the ejecta in this location has
indeed recently encountered the remnant's reverse shock front. 

\citet{anderson94} predict that ejecta knots which are strongly decelerated
by the reverse shock will show both a marked increase in their thermal
emission (like that seen here), coupled with an increase in synchrotron emission.
We have compared 6 cm VLA radio observations ($\sim$ $0 \farcs 4$ 
resolution) taken in April 2000 and again in April 2001 and note a 
qualitative increase in the
radio emission around Region 2 \citep{delaney06} which is comparable to 
the increase in X-ray emission. This increase in the radio and 
X-ray flux is consistent with the interpretation of \citet{anderson94}.
However, we do not find any corresponding
changes in the radio emission for the other five regions we present here,
contrary to what is expected if we interpret these flux changes in
the context of \citet{anderson94}. 
In particular, Region 1, which like Region 2, also shows a change in 
$kT$ (from 0.8 -- $\sim$ 1.1 keV) between 2000 and 2004 , does not exhibit 
a corresponding increase in the 6 cm radio emission \citep{delaney06}.

On the other hand, model fits for Region 4, a large diffuse patch along the 
eastern rim and which showed about the same level of X-ray flux changes as 
seen for 
Region 1, exhibited nearly a constant electron temperature over the four year 
period.  Moreover, Region 3 (Fig.~\ref{fig:reg34}, left) did not exhibit the 
same sort of strong, monotonic $kT$ changes as the other three regions. 
However, based on the large change in the modeled $kT$ between 2000 and 2002 
(Table~\ref{tab:mfits}) from $\sim$ 0.8--1.5 keV, along with the emergence of 
a broadening near helium-like sulfur (tentatively identified here as Si 
Ly$\beta$) suggests that the ejecta here may be undergoing the same sort of 
rapid ionization seen in Region 2. Another possible spectral change seen for 
Region 3 is the apparent broadening of the silicon lines in the 2002 and 2004 
spectra.  Combined with the increase in $kT$, this suggests that the ejecta is 
in the process of ionizing up to the hydrogen-like state for silicon.

From its spectrum, Region 4 (Fig.~\ref{fig:reg34}; right) appears to be nearly
in ionization equilibrium, and as noted above (Table~\ref {tab:mfits}) the
fitted temperature stays nearly constant between 2000 and 2004. Modeling
suggests that its spectrum can be equally well modeled by either a  NEI plasma
with an ionization age $\sim$ 10$^{13}$ cm$^{-3}$ s, or by a collisional
ionization model, such as a Raymond--Smith plasma \citep{raymond77}.  The high
fitted ionization age is puzzling because it implies electron densities of
$\sim$ 10$^{3}$ cm$^{-3}$; densities which are as high as those seen in the
optical knots. However, visual inspection of the spectrum for Region 4
(Fig.~\ref{fig:reg34}; right) shows that it is clearly different from regions
which do not exhibit ionization equilibrium (i.e., Regions 1 and 3 for
instance). 

\citet{morse04} proposed that Cas~A's ejecta consists of cool, optically bright
emission knots embedded in a warm, tenuous ejecta medium which is responsible
for the observed X-ray emission. This general description is supported by our
model fits (Table~\ref{tab:mfits}). Were the ejecta completely uniform, then a
single temperature, single ionization age would adequately describe the spectra
in all regions at each epochs. However, since we see that the fitted
temperature rises in Regions 1--3, the rate at which the ejecta is being
shocked in these regions is not constant due to density inhomogeneities in the
preshock material.

While the four regions which showed increases in intensity appear to be
associated with ejecta which has recently crossed the reverse shock,  Regions 5
and 6 show flux decreases and therefore do not indicate reverse shock
locations.  The interior projected thin filament seen in Region 5 has a
non-thermal spectrum and is likely associated with the forward and  not the
reverse shock front \citep{delaney04}.  The observed changes in its  brightness
probably reflects changes in path length along interconnecting  tangents of the
wave front of the forward shock as it interacts with the local  circumstellar
medium on the near-facing side of the remnant. 

In contrast,  emission  seen in
Region 6 (Fig.~\ref{fig:reg56}, right) which also showed a decrease in
brightness, may represent ejecta cooling in the post-shock flow  following
reverse shock passage. Its projected location in the remnant is consistent with
this identification.  While it lies at approximately the same radial  distance
from the center of expansion as the brightening Regions 1--4, it lies just
behind (eastward) a patch of optically bright emission ejecta along the
remnant's southwestern limb. If this feature is indeed fading due to cooling,
this would imply it was shocked much earlier than in the other regions in the
area and would be consistent with its location inside the steady X-ray and
optical  emission features found farther out from the remnant's center.

\subsection{Scale Length of X-ray Emitting Knots}

Rapid intensity changes in small regions throughout the remnant offer some 
insight into the structure of the lower density component of Cas~A's SN 
ejecta. Abrupt increases (or decreases) in the observed count rate strongly 
suggest that the X-ray emitting ejecta is non-uniform on fairly small scales. 
This conclusion has long been clear from optical observations 
\citep{Fesenetal01} but not from  X-ray observations which are sensitive to a 
much lower density, higher temperature component of the ejecta but which likely 
dominate the remnant's overall mass budget \citep{Vink96,willingale03,
laming03}.  The velocity of the freely expanding ejecta clumps at the 
specific regions we investigated is not known, but a rough estimate can be 
made based on the distance of the reverse shock from the center of expansion. 
The reverse shock is located $r_{rs}$ $\sim$ 95$\arcsec$ ($\simeq$ 1.6 pc at 
a distance of 3.4 kpc) from the center of expansion \citep{gotthelf01} and 
unshocked, freely expanding ejecta at this distance have been expanding for 
$t_{age}$ $\simeq$ 330 yr.  Therefore, we estimate that the unshocked ejecta 
velocity is no more than $v_{ej}$ $\approx$ $r_{rs}/t_{age}$ $\approx$ 5000 
km s$^{-1}$.

The observed intensity for the small features we studied show significant 
changes on time scales of $\Delta t$ $\sim$ 2--4 yrs. This suggests a scale 
length for spatial variations in the ejecta no larger than the distance 
covered by a $\sim$ 3000 km s$^{-1}$ reverse shock moving through $\sim$ 5000 
km s$^{-1}$ expanding ejecta over the course of a few years; i.e., $\Delta 
t \times v_{ej}$ $\approx$ $0.02 - 0.03$ pc or about 1--2 arcsecond at 3.4 
kpc. This scale length is consistent with the actual sizes of the variable 
flux features we see in the remnant. In contrast, the fine-scale structure 
of optical knots in Cas~A are of order $\sim 0\farcs2 - 0\farcs3$ in size 
\citep{Fesenetal01}, roughly an order of magnitude smaller than the spatial 
scale for variations seen in the clumpy component of the X-ray bright ejecta.

\section{Conclusions}

With the exception of the rapidly evolving remnant associated with  SN~1987A
\citep{Burrows00,Park04}, no remnant has been previously reported to  show
significant flux changes across portions of its X-ray emitting  structure.
This is due in part to the limited number of high-resolution images available
on young SNRs covering a significant time span, and the relatively small number
of very young Galactic or LMC/SMC SNRs known.

Inspection of X-ray data taken of the Cassiopeia A supernova remnant  with {\it
Chandra} ACIS-S3 in 2000, 2002, and 2004 reveals the presence of  several small
scale features ($\leq 10\arcsec$) which exhibit significant intensity  changes
over this 4 year time frame. Here we described six such features four  of which
had count rate increases from $\sim$ 10\% to over 90\%, while two  others
showed decreases of $\sim$ 30\% -- 40\%. Whereas the majority of the six
variable flux features exhibited no gross spectral changes in the 1--4 keV
energy band during the four year period, one region which had the greatest
increase in brightness also showed hints of slight emission line changes as
well as an increase in its 6 cm radio emission.
Spectral fits using non-equilibrium ionization, metal-rich plasma models
indicate increased or decreased electron temperatures for 
features showing increasing or decreasing count rates, respectively. 

A fading emission
filament  projected near the remnant's interior (Region 5) showed a completely
nonthermal spectrum  and thus may be associated with the forward shock front.
It decreasing  intensity may be the result of changing path lengths along the
edges of a wavy forward  shock as it interacts with the local, inhomogeneous
circumstellar medium.  Based  on the observed count rate changes and the
assumption of freely expanding 5000 km s$^ {-1}$ ejecta at the reverse shock
front, we estimate the  unshocked ejecta to have spatial scale variations of
$0.02 - 0.03$ pc,  consistent with the X-ray emitting ejecta belonging to a
more diffuse component of the supernova ejecta.

In general, bright X-ray and optical knots are not coincident. This is not
unexpected given the differences in their temperatures and densities.
Moreover, the sizes of the X-ray knots ($\sim1''-5''$)
appear to be an order of magnitude larger than the optical knots
($\sim0 \farcs2-0 \farcs3$; \citealt{Fesenetal01}).  However, the remnant's optical knots
can also appear and brighten on timescales as short as a few years, not unlike
the timescales we observed for these X-ray knots.  Given the newly observed short term
variability seen here in the X-ray, future X-ray and optical observations
of selected knots and filaments might provide valuable information regarding
the dynamical and radiative evolution of supernova ejecta knots over a range
of densities, sizes, and postshock temperatures.

Cassiopeia A is currently the youngest and one of the most studied Galactic
SNRs. The small yet noticeable changes we found combined with future
X-ray observations of this bright X-ray remnant may help shed light not only
on the remnant's near-term X-ray evolution, but also help us better
understand the fine-scale structure, dynamics and chemistry of debris from
core-collapse supernovae in general.

\acknowledgements
The authors wish to thank Paul Plucinsky for helpful advice regarding
the accuracy of the ACIS response matrices and Tracey Delaney for  
discussions
regarding the emergence of radio knots coincident with the X-ray knots. We
would also like to thank the referee, Parviz Ghavamian, for several 
valuable comments and suggestions.
This work was partially supported by the
Chandra X-ray Observatory through the Archival Research Award AR4-5005X.

\begin{deluxetable}{cccccccc}
\tabletypesize{\scriptsize}
\tablecolumns{8}
\tablewidth{0pc}
\tablecaption{ACIS-S3 Count Rates}
\tablehead{
\colhead{Region} & \colhead{RA (J2000)} & \colhead{Dec (J2000)} &
\colhead{Extraction} &
\multicolumn{3}{c}{Normalized Count Rate\tablenotemark{a}} &
\colhead{Count Rate Change}\\
\colhead{} & \colhead{(h:m:s.s)} & \colhead{(\degr~ ~\arcmin~ ~
\arcsec)} &
\colhead{Region} & \colhead{(2000.1)} & \colhead{(2002.1)} &
\colhead{(2004.1\tablenotemark{b}~)} & \colhead{($\Delta_{2004-2000}
$)} }
\startdata
1  & 23:23:34.0 & +58:49:56.2 & $5'' \times 4''$ & 1.84$\pm$0.02 &
2.04$\pm$0.01 & 2.09$\pm$0.01 & $+14$\% \\
2  & 23:23:24.4 & +58:47:13.2 & $3'' \times 3''$ & 0.48$\pm$0.02 &
0.57$\pm$0.01 & 0.93$\pm$0.01 & $+93$\% \\
3  & 23:23:13.4 & +58:48:10.0 & $5'' \times 3''$ & 0.93$\pm$0.01 &
1.28$\pm$0.02 & 1.10$\pm$0.02 & $+18$\% \\
4  & 23:23:11.2 & +58:48:53.2 & $5'' \times 5''$ & 2.06$\pm$0.02 &
2.24$\pm$0.02 & 2.51$\pm$0.01 & $+22$\% \\
5  & 23:23:26.9 & +58:48:19.0 & $3'' \times 10''$& 1.56$\pm$0.02 &
1.21$\pm$0.02 & 0.95$\pm$0.02 & $-40$\% \\
6  & 23:23:18.0 & +58:48:37.0 & $4'' \times 4''$ & 1.04$\pm$0.01 &
0.83$\pm$0.01 & 0.71$\pm$0.02 & $-31$\% \\
\enddata
\tablenotetext{a}{Values listed are detected counts s$^{-1}$
normalized to Cas A's X-ray point source rate
                    of 0.12$\pm$0.01 cts s$^{-1}$.}
\tablenotetext{b}{Epoch 2004 count rates are calculated from {\it
Chandra} Observation ID
5196, which has the same roll angle as the 2000.1 and 2002.1
observations.}
\label{tab:rates}
\end{deluxetable}

\begin{deluxetable}{cccc}
\renewcommand{\arraystretch}{1.5}
\tablecolumns{4}
\tablewidth{0pc}
\tablecaption{Spectral Fit Temperature Variations}
\tablehead{
\colhead{Region} & \multicolumn{3}{c}{$kT$ (keV)} \\
\colhead{} & \colhead{2000.1} & \colhead{2002.1} & \colhead{2004.1}}
\startdata
1 & 0.76$_{-0.03}^{+0.02}$ & 0.86$_{-0.06}^{+0.09}$ & 1.05$_{-0.10}^
{+0.12}$ \\
2 & \nodata & 1.09$_{-0.28}^{+0.32}$ & 1.46$_{-0.12}^{+0.17}$ \\
3 & 0.84$_{-0.05}^{+0.06}$ & 1.43$_{-0.10}^{+0.07}$ & 1.45$_{-0.12}^
{+0.08}$ \\
4 & 0.94$_{-0.03}^{+0.02}$ & 0.92$_{-0.02}^{+0.04}$ & 0.94$_{-0.05}^
{+0.02}$ \\
6 & 0.71$_{-0.03}^{+0.04}$ & 0.63$_{-0.02}^{+0.02}$ & 0.55$_{-0.04}^
{+0.02}$ \\
\enddata
\label{tab:mfits}
\end{deluxetable}

\clearpage

\end{document}